\documentclass[12pt]{article}
\usepackage[normalem]{ulem}
\usepackage[mathscr]{eucal}
\usepackage[version=3]{mhchem}
\usepackage{stmaryrd}
\usepackage{hyperref}

\usepackage{epsfig,latexsym,amsfonts,amsmath,amsthm,amssymb,amsbsy,multirow,slashed,wasysym,textcomp,dsfont,comment,mathtools,cancel,cite,datetime,appendix,BOONDOX-calo}
\usepackage{tocloft}
\usepackage{bbold}
\usepackage{tikz}
\usetikzlibrary{decorations.pathreplacing,calligraphy}
\usetikzlibrary{backgrounds}
\usetikzlibrary{patterns}
\usetikzlibrary{arrows.meta}
\tikzstyle{bag} = [align=center]
\usetikzlibrary{decorations.pathmorphing}
\def\bea{\begin{eqnarray}}
\def\eea{\end{eqnarray}}

\usepackage{hyperref}
\usepackage{subcaption}

\usepackage{array}
\usepackage{booktabs}

\newcommand{\badat}{\begin{alignedat}}
\newcommand{\eadat}{\end{alignedat}}

\def\be{\begin{equation}}
\def\ee{\end{equation}}
\def\ba{\begin{aligned}}
\def\ea{\end{aligned}}

\usepackage{xcolor}

\newcommand{\pink}[1]{\textcolor{\pink}{#1}}

\definecolor{dblue}{rgb}{0.2,0.50,0.80}

\DeclareFontFamily{OT1}{pzc}{}
\DeclareFontShape{OT1}{pzc}{m}{it}{<-> s * [1.10] pzcmi7t}{}
\DeclareMathAlphabet{\mathpzc}{OT1}{pzc}{m}{it}

\definecolor{vert}{rgb}{0.1367 0.543 0.1367}

\numberwithin{equation}{section} % equation numbers follow sections

\begin{document}

 \begin{titlepage}
  \thispagestyle{empty}
  \begin{flushright}
%%%% report number
  \end{flushright}
  \bigskip
  \begin{center}

        \baselineskip=13pt {\LARGE {
       An Exact Black Hole Scattering Amplitude
       }}
   
      \vskip1cm 

   \centerline{ 
   {Alfredo Guevara} ${}^{{}\spadesuit,{}\heartsuit{}}$ 
   {Uri Kol} ${}^\diamondsuit{}$ and {Huy Tran}  ${}^{\clubsuit}$ 
}

\bigskip\bigskip

\centerline{\em${}^\clubsuit$ 
\it Department of Physics, Harvard University, Cambridge, MA 02138}
\bigskip

%\centerline{\em${}^\clubsuit$ 
%\it Black Hole Initiative, Harvard University, Cambridge, MA 02138
%}

%\bigskip

\centerline{\em${}^\spadesuit$ 
\it Society of Fellows, Harvard University, Cambridge, MA 02138
}

\bigskip

\centerline{\em${}^\heartsuit$ School of Natural Sciences, Institute for Advanced Study, Princeton, NJ 08540 USA
}
 \bigskip
 
 \centerline{\em${}^\diamondsuit$ 
\it Center for Mathematical Sciences and Applications, Harvard University, MA 02138}

\bigskip\bigskip

\end{center}

\begin{abstract}
 \noindent 
General Relativity famously predicts precession of orbital motions
in the Schwarzschild metric. In this paper we show that by adding a
NUT charge $N=iM$ the precession vanishes to all orders in $G$ even for rotating black holes. Moreover,
we conjecture a generalization of the eikonal formula and show that the classical integrable
trajectories determine the full quantum amplitude for this black hole,
by means of exponentiation of the Post-Minkowskian radial action. Several consequences of integrability in self-dual gravity are discussed.
  
\end{abstract}

\end{titlepage}

\tableofcontents

\section{Introduction}

As the precision of strong gravity tests continues to increase, classical black hole dynamics in asymptotically flat gravity is one of the leading technical challenges on its theoretical frontier. A great deal of recent progress stems from mathematical structures emerging in quantum theory, such as scattering amplitudes, that can be nevertheless implemented to tract the classical behaviour of gravitating objects. In this direction, the end goal is to encode black hole dynamics into gauge invariant quantities whose systematic computation is well under control via modern \textit{on-shell} methods.

As a matter of fact, to explicitly see the advantage of these methods, it is desirable to find solvable instances of the above challenges where dynamics is completely under control. This is the purpose of this work. More precisely, our motivation is to analyze the wave equation
\begin{equation}
\label{eq:wvq}
    g^{\mu \nu}\nabla_{\mu}\nabla_{\nu}\psi = \frac{m^2}{\hbar^2} \psi 
\end{equation}
on a background corresponding to a particular rotating black hole. The solution of the wave equation is directly connected to the black hole Green function and to its thermal spectrum. In fact, even though the frequency domain of this equation is certainly well studied in BH backgrounds, featuring the celebrated quasi-normal modes (QNM), much less is known about its analytic properties in position space.\footnote{Some notable exceptions to this ignorance is the property of the massless Green function of having poles in the geodesic distance \cite{DEWITT1960220}.}

For type D spacetimes, separation of variables in \eqref{eq:wvq} is possible and one finds a well-posed problem in terms of ODEs. However, for spinning-charged backgrounds the radial problem is the (in)famous confluent Heun equation which is hard to tract. Some simplifications emerge in the geodesic (i.e. eikonal) limit, which however still leads to elliptic integrals to be evaluated numerically. Because of this, it becomes important to find solvable backgrounds where the above can be studied analytically. An interesting candidate in this direction are BPS instantons, also known as KK monopoles, which arise in supersymmetric theories. In particular the two-point function in these backgrounds was derived using the solution of the Killing spinor equation \cite{Gibbons:1978tef,Page:1979ga}.

Recently the KK monopoles solutions have been reinterpreted as self-dual black holes \cite{Crawley:2023brz} (see also \cite{Easson:2023ytf}), namely solutions of the self-dual sector of GR which make sense in $(2,2)$ signature. These solutions are static and spherically symmetric. Surprisingly, however, they are smoothly connected to astrophysical Kerr black holes, carrying spin $a$ and mass $M$, by the inclusion of a NUT charge $N=\pm i M$ in Lorentzian signature. The spin parameter in this description can be removed, linking different aspects of static and rotating solutions. These black holes have revealed exciting properties closely related to integrability of self-dual GR, in particular making possible the computation of amplitudes \cite{Guevara:2023wlr,Adamo:2023fbj,Adamo:2024xpc} (see also \cite{Guevara:2021yud,Kim:2024dxo}).

In this paper, we continue our exploration of the two point-function of this black hole with two main purposes. The first one is to show explicitly that at the self-dual (SD) point the two-point function, namely a solution to \eqref{eq:wvq}, is controlled by classical geodesics via a localized path integral. To achieve this, we solve the particle motion in static and rotating scenarios in the presence of a NUT charge and show that integrability emerges at the self-dual point. Then we argue that scattering solutions are obtained by exponentiating unbounded classical trajectories. This is closely related to a leading WKB approximation, but at the self-dual point, it becomes exact via a new eikonal-type formula:
\begin{equation}
f(k)=\frac{p}{2\pi}\int d^{2}be^{ik\cdot b+i2q\phi}e^{i\chi(b)} \label{eq:intrs}
\end{equation}
Here $\chi (b)$ is the eikonal phase with impact parameter $b$, obtained purely from the geodesic analysis. This generalizes the well known eikonal formula by including a phase proportional to $q := 2 N \omega$ to be integrated over. 

The second main purpose of this work is to understand the implications of the above beyond the self-dual point, namely into full GR and its two body problem. In \cite{Guevara:2023wlr} we formulated the self-dual problem as a Hydrogen atom system in quantum mechanics, and proposed that corrections can be accounted in perturbation theory similarly to the Zeeman effect. Here we will revisit corrections in light of the exact WKB approximation. \footnote{In forthcoming work \cite{upcoming:2025} we will recast such as deviations from extremality in GR.} For instance, the degeneracy of the spectrum in the hydrogen atom is a consequence of closed bounded orbits in its classical counterpart. By computing the periastron precession $\Delta \Phi$ in backgrounds slightly away from the self-dual point, the degeneracy lift can be accounted in terms of an average angular velocity $\Delta \Phi/\Delta T$. The precession $\Delta \Phi$ is a gauge invariant observable closely related to the phase $\chi(b)$ entering the amplitude \eqref{eq:intrs} (see \cite{Kalin:2019inp}). We will show that it is in fact determined by an analytic discontinuity of $\chi(b)$ beyond the self-dual point.

The plan for this work is as follows. In Section \ref{sec:geo} we analyze the classical orbits of the SD black hole and make an analogy with the orbits in a monopole background, due to the corrections of the classical angular problem. In section \ref{sec:wkb} we discuss how these orbits first approximate the quantum amplitude on this background via the usual saddle point WKB expansion. Based on this we then introduce the new eikonal exact formula in section \ref{sec:pmf}. In section \ref{sec:comps} we also present an alternative route where the amplitude is computed from a particular complex structure. Section \ref{sec:contd} is devoted to extensions and future directions, particularly the massless S-Matrix defined via analytic continuation, and general helicity scattering. 

\section{Classical Geodesics}\label{sec:geo}

We start by considering the general Taub-NUT solution corresponding to Schwarzschild metric endowed with a non-zero NUT charge $N$. This background is interesting for several reasons, for instance in (2,2) signature it allows for scattering solutions and the conical deficit can be removed. In principle we consider the spinless case ($a=0$), but as we move forward, we will focus on the self-dual point for which $a$ is irrelevant.

The goal is to extract the information corresponding to classical trajectories in a gauge invariant manner. We will carry
the analysis first in Euclidean signature and analytically continue later. The Taub-NUT metric is given by
\begin{equation}
ds^{2}=\frac{\Delta(r)}{r^{2}-N^{2}}\left(dt+2N(\zeta-\cos\theta)d\phi\right)^{2}+(r^{2}-N^{2})\left(dr_{*}^{2}+d\theta^{2}+\sin^{2}\theta d\phi^{2}\right),\label{eq:tnsx}
\end{equation}
where $\Delta=N^{2}+r(r-2M)$ and $dr_{*}=\frac{dr}{\sqrt{\Delta(r)}}$. Our first strategy is
to compute geodesics in this metric using Hamilton-Jacobi theory,
which can then be promoted to a WKB approximation. We will see that
this approach is almost exact for the quantum amplitude, but misses a phase. The
exact amplitude will be derived from a high-energy exponentiation
in section \ref{sec:pmf}.

Let us start with the on-shell action constraint, a.k.a. the Hamilton Jacobi equation,
\begin{equation}
g^{\mu\nu}\partial_{\mu}S\partial_{\nu}S=m^{2}\,.\label{eq:hjeq}
\end{equation}
We will use the following splitting: 
\begin{align}
S & =\omega t+S_{\theta,\phi}(J,J_{z},q)+S_{r}(M,N,J,\omega)\nonumber \\
 & =\omega t+\int(p_{\theta}d\theta+p_{\phi}d\phi)+\int p_{r}dr_{*}\label{eq:refas}
\end{align}
The integration constants are defined as usual, i.e. we introduce

\begin{equation}
p_{r}=\frac{\partial S_{r}}{\partial r_{*}}\,\,,p_{\theta}=\frac{\partial S_{\theta,\phi}}{\partial\theta}\,,p_{\phi}=\frac{\partial S_{\theta,\phi}}{\partial\phi}=J_{z}+\zeta q
\end{equation}
Assuming as usual that $p_\theta$ and $p_r$ are functions of $\theta$ and $r$ respectively, equation (\ref{eq:hjeq}) takes a separable form. Further defining $q=2N\omega$, $p^{2}=m^{2}-\omega^{2}$, and $J^2$ as the separation constant, it splits into:
\begin{align}
p_{\theta}^{2}+\frac{(J_{z}+q\cos\theta)^{2}}{\sin^{2}\theta} & =J^{2}-q^{2}\,\,\,\label{eq:asdw}\\
\frac{p_{r}^{2}}{r^{2}-N^{2}}= &~ p^{2}-\frac{J^{2}}{r^{2}-N^{2}}-\frac{2M\omega^{2}}{r+N}-\underbrace{2\frac{M-N}{r^{2}-N^{2}}\omega^2 \left(N+\frac{2r^{2}(M+N)}{\Delta}\right)}_{\mathcal{O}(G^2)}\,.\nonumber 
\end{align}
We see that the Post-Newtonian terms $\mathcal{O}(G^{2})$ vanish
at the self dual point $M=N$. As will be reviewed, perihelion precession is indeed a $\mathcal{O}(G^2)$ effect, and thus its absence in the radial action is a manifestation of integrability of the theory.

Let us now integrate the radial component
\begin{align}
S_{r} & =\int_{r_{0}}^{r}\frac{p_{r}}{\sqrt{\Delta(r)}}dr=\int_{r_{0}}^{r}\sqrt{\frac{p^{2}(r^{2}-N^{2})-J^{2}-2M\omega^{2}(r-N)+\mathcal{O}(M-N)}{\Delta(r)}}dr\nonumber \\
 & \stackrel{r\to r+M}{=}\int_{r_{0}}^{r}\sqrt{p^{2}(1+\frac{2M}{r})-\frac{J^{2}}{r^{2}}-\frac{2M\omega^{2}}{r}}dr+\mathcal{O}(M-N)\nonumber \\
 & = \left. p_{r}+\mathcal{M}p\tanh^{-1}\left[\frac{p}{p_{r}}(r+\mathcal{M})\right]-J\arctan\left[\frac{p^{2}r\mathcal{M}-J^{2}}{Jp_{r}}\right] \right|^{r}_{r_0}+\mathcal{O}(M-N)\label{eq:rct}
\end{align}
where $\mathcal{M}=M(1-\omega^{2}/p^{2})$ (note that in the massless
limit $i\mathcal{M}p=2M\omega$, this will be our Post-Minkowskian
expansion parameter later). We note that
\begin{equation}
    p_{r}=\sqrt{p^{2}r^{2}+2p^{2}r\mathcal{M}-J^{2}} \label{eq:sam1}
\end{equation}
at the SD point. At the SD point, the radial action matches the Coulomb problem with a Newtonian effective potential \footnote{The impetus formula of \cite{Kalin:2019inp}, argues that $V(r)$ is the
3d Fourier transform of the (tree-level) amplitude. Analogously the
$\mathcal{O}(G^{2})$ terms in (\ref{eq:asdw}) should match the higher-loop
amplitude and thus vanish in the self-dual point. More on this in
section 3.}

\begin{equation}
V(r)=\frac{2\mathcal{M}p^{2}}{r}\,.
\end{equation}
Note that although
the potential is linear in $\mathcal{M}\sim G$ the radial action
contains all orders in $G$ due to the presence of the square root. 

To solve the angular action let us adjust the axis so that $\theta$
is a scattering angle. This is achieved when the particle comes from
the z-axis, thus $J_{z}=q,J_{x}=L$. Then (\ref{eq:asdw}) becomes
\begin{equation}
p_{\theta}=\sqrt{J^{2}-q^{2}\csc^{2}\theta/2}
\end{equation}
and
\begin{align}
S_{\theta,\phi}(J,q) & =\int p_{\theta}d\theta+q(1+\zeta)d\phi\nonumber \\
 & =\int\sqrt{J^{2}-q^{2}\csc^{2}\theta/2}d\theta\nonumber \\
 & =2q\arcsin\left(\frac{q\cot\theta/2}{\sqrt{J^{2}-q^{2}}}\right)-2J\arcsin\left(\frac{J\cos\theta/2}{\sqrt{J^{2}-q^{2}}}\right)+c \label{eq:angact}
\end{align}
where we have set $\zeta=-1$ to ignore the $\phi$ dependence for
convenience. Note that the turning points of $S_{\theta,\phi}$ are located at $\sin\theta/2=\pm q/J$. Bohr-Sommerfield quantization then leads
to 
\begin{equation}
    2q-2J=n \in \mathbb{Z}\,. \label{eq:angqt}
\end{equation}
It is familiar from monopole scattering, and it will be clear in the next section, that in the presence of a NUT charge the particle motion is restricted to a cone. The angle swiped in the cone section is indeed
\begin{equation}
\frac{\beta}{2}:=-\frac{1}{2}\frac{\partial S_{\theta}}{\partial J}=\arcsin\frac{J\cos\theta/2}{\sqrt{J^{2}-q^{2}}} \label{eq:sxa}
\end{equation}
As $q\to 0$ the cone degenarates to a plane and $\beta \to \theta - \pi $ becomes the classical deflection angle. Since the angular action can be evaluated exactly for any $M,N$ the relation between $\beta$ and $\theta$ given by \eqref{eq:sxa} is exact, and we regard $\beta$ itself as the observable. It is fixed
by the Hamilton-Jacobi equation:%\footnote{We can write $S_{\theta,\phi}=2q\arcsin\left(\frac{\cot\theta/2}{\tan\chi/2}\right)-2J\arcsin\left(\frac{\cos\theta/2}{\sin\chi/2}\right)$
%where $\cos\chi/2=q/J$. Then $\frac{\partial S_{\theta,\phi}}{\partial\chi}=0$.
%This is equivalent to

%\begin{align*}
%\frac{\partial S_{\theta,\phi}}{\partial J} & =-\beta(\theta,J)\\
%\frac{\partial\beta}{\partial J} & =\cos\frac{\chi}{2}\frac{\partial\alpha}{\partial J}
%\end{align*}
%This is also true for the radial action $\frac{\partial S_{r}}{\partial\chi}=0$
%at least in the SD point.}
%
\begin{equation}
0=2\frac{\partial S_{r}}{\partial J}+\frac{\partial S_{\theta,\phi}}{\partial J}
\end{equation}
where the factor of 2 accounts for the return trajectory
\begin{align}
\frac{\partial S_{r}}{\partial J}=&\frac{\pi}{2}-\arctan\left[\frac{p^{2}r\mathcal{M}-J^{2}}{Jp_{r}}\right] \\
\stackrel{r\to \infty}{=}& \frac{\pi}{2}-\arctan\left(\frac{p\mathcal{M}}{J}\right)
\end{align}
We have taken $r\to\infty$ to get the scattering angle. Further introducing the impact parameter $b=J/p$ we see that the above expression resums all Post-Minkowskian orders in $\frac{\mathcal{M}}{b}$ as anticipated. However, corrections supressed by $\frac{M-N}{b}$ are obtained by expanding the integrand of \eqref{eq:rct}.\footnote{The lower limit of integration shall also be expanded by it does not contribute to the leading correction quoted here.} To illustrate this, we point out that the first correction reads
\begin{align}
S_r^{(1)}=& \frac{(l-M)M}{2J^5}\left(\frac{J p_r}{r^2}(J^4 +J^2q^2 + r \chi_J^{(0)})+\chi_J^{(1)} \arctan\left[\frac{p^{2}r\mathcal{M}-J^{2}}{Jp_{r}}\right] \right)^{r\to \infty}_{r_0}\\
=&  \frac{(l-M)M}{2J^5}\left(J p \chi_J^{(0)}+ \chi_J^{(1)} \arctan\left[\frac{J}{p\mathcal{M}}\right] \right)\label{eq:srby}
\end{align}
with
\begin{align}
    \chi_{J}^{(0)}=&3 q^2 \mathcal{M}p^2{-}J^2 \mathcal{M}p^2{+}16 J^2 M \omega^2\\
    \chi_{J}^{(1)}=& 3 J^4 (p^2 - 4 \omega^2) - 12 M^4 (-p^2 \omega + \omega^3)^2 + J^2 M^2 (p^4 - 22 p^2 \omega^2 + 17 \omega^4)
\end{align}
At large $J\sim \mathcal{M}p$ the correction to the scattering angle goes as $\frac{\partial S_r^{(1)}}{\partial J} \to \frac{3\pi (N- M) M (p^2 - 4\omega^2)}{4J^2}$ and thus it is supressed by $\frac{(N-M)}{b}$. The splitting in polynomials $\chi_{J}^{(i)}$ is motivated by the trascendentality weight in $J$. Indeed we will discover that only the trascendental component $\chi_{J}^{(1)}$ enters in the perihelion precession computation. This is because it features a discontinuity in complex $J$ plane. It would be interesting to examine this fact at higher orders in the expansion, in conjunction with the prescription proposed in \cite{Kalin:2019inp}.

\subsection{Precession constraint}\label{sec:prc}

Using the above results, we can further analyze the precession of
an orbit as a test of integrability. This corresponds to bounded dynamics
as opposed to the scattering setup.

We will perform a standard rotation to set $J_{z}=J$ in (\ref{eq:asdw}).
In that case the radial equation is unmodified but the angular system
becomes
\begin{equation}
p_{\theta}^{2}+\frac{(J\cos\theta+q)^{2}}{\sin^{2}\theta}=0\,\,\,,J_{z}=J \label{eq:tfind}
\end{equation}
The two terms being positive must vanish separately. This has the
meaning of setting $J_{x}=J_{y}=0$, and leads to the particle orbiting
the cone
\begin{align*}
p_{\theta} & =0\\
\cos\theta & =-q/J
\end{align*}
From the first equation the angular action now integrates as $S_{\theta,\phi}=J\Delta\phi$,
and the HJ equation now gives
\begin{equation}
\frac{\partial S_{r}}{\partial J}=-\Delta\phi\label{eq:scm}
\end{equation}
When evaluating $S_{r}$ between the two roots of $p_{r}$, say $r_{0}'<r_{0}$,
we see that the corresponding arc $\Delta\phi$ is the precession
angle modulo $\pi$. Our task is then to evaluate
\begin{equation}
S_{r}^{\textrm{bound}}=\int_{r_{0}'}^{r_{0}}\frac{p_{r}}{\sqrt{\Delta(r)}}dr\,,\quad p(r_0)=p(r'_0)=0\,.
\end{equation}
We can evaluate this integral at any order in $(N-M)/J$. At leading order, namely $N=M$, we note that $p_r$ has two branch points so that the integral is performed along the cut. In fact this holds when expanding the integrand to any order, assuming uniform convergence. Further branch points of $\frac{p_r}{\sqrt{\Delta(r)}}$ are deformed to $r\to 0$ as $N\to M$ and the integral can be evaluated, at a given order, by its residues at $r=0,\infty$. At strict $M=N$ we find
\begin{equation}
S_{r}^{\textrm{bound}}=\int_{r_{0}'}^{r_{0}}\frac{p_{r}}{\sqrt{\Delta(r)}}dr=(i\mathcal{M}p-J)\pi\,,
\end{equation}
where the first (second) term is the residue at $r=\infty$ ($r=0$). 
This confirms through (\ref{eq:scm}), that there is no precession at
the self-dual point.

In general, only the residue at $r=0$ depens on $J$ and contributes to the precession. It can be shown that this residue is precisely computing the discontinuity in $J$ of the scattering result, e.g. \eqref{eq:srby}. Indeed, the first correction beyond the self-dual point reads
\begin{align}
S_r^{\textrm{bounded},(1)}=& \frac{(l-M)M\pi}{2J^5} \chi_J^{(1)} 
\end{align}
leading to
\begin{equation}
    \Delta \phi =\pi \left(1+\frac{(l-M)M}{2}\frac{\partial}{\partial J} (J^{-5} \chi_J^{(1)}) \ldots \right)
\end{equation}

A corollary of the above is that the radial part of the SD black
hole can be mapped to the hydrogen atom \cite{Guevara:2023wlr}. For instance, the WKB spectrum
of bound states is given by 
\begin{equation}
i\pi n=S_{r}^{\textrm{bound}}(\omega,J)+\pi J
\end{equation}
where we can solve $\omega=\omega(n,J)$ perturbatively and find $\omega= \frac{n}{2M}+\mathcal{O}(M-N)$. In general, the WKB frequencies satisfy
\begin{equation}
    \frac{\partial \omega(n,J)}{\partial J}=\Delta \phi - \pi
\end{equation}
and so the absence of precession indicates degeneracy of the spectrum. As in the hydrogen atom, this degeneracy is lifted in perturbation theory.

\subsection{Kerr Black Holes}\label{sec:kerr}

It is easy to generalize the discussion for non-zero spin $a$. The
addition of spin is encapsulated by the Kerr Taub-NUT metric,
\begin{align}
ds^{2} & =f(dt+\Omega d\phi)^{2}+\frac{R^{2}}{\Delta}d\rho^{2}+R^{2}(d\theta^{2}+\Sigma^{2}\sin^{2}\theta d\phi^{2})\nonumber \\
\Delta(\rho) & =\rho^{2}-2M\rho-a^{2}+N^{2}\nonumber \\
R^{2} & =\rho^{2}-(N-a\cos\theta)^{2}\nonumber \\
f & =1-\frac{2Mr-2N(N-a\cos\theta)}{R^{2}}\nonumber \\
\Omega & =-2N\cos\theta-a(1-f^{-1})\sin^{2}\theta\,,\label{eq:sd1xa}
\end{align}
whose $N\to0$ limit is the standard Kerr metric, while $a\to0$ is
the Taub-NUT solution. Imposing the HJ equation (\ref{eq:hjeq}) in
this background leads to
\begin{equation}
S=\omega t+p_{\phi}\phi+\int p_{x}dx+\int p_{\rho}d\rho\label{eq:spinxq}
\end{equation}
(setting $x=\cos\theta$ as usual), with
\begin{align}
p_{x} & =\frac{\sqrt{\Theta(x)}}{1-x^{2}}\,,\,\Theta(x):=(1-x^{2})\left[Q-p_{\phi}^{2}+2Nax(m^{2}-2\omega^{2})+a^{2}(\omega^{2}-m^{2})(x^{2}-1)\right]-(q+xp_{\phi})^{2}\nonumber \\
p_{\rho} & =\frac{\sqrt{\mathcal{P}(\rho)}}{\Delta(\rho)}\,,\,\mathcal{P}(\rho):=\Delta(\rho)\left[Q-q^{2}+2ap_{\phi}\omega-(\rho^{2}-a^{2}-N^{2})m^{2}\right]-(\omega(\rho^{2}-a^{2}-N^{2})-ap_{\phi})^{2}\label{eq:pxprho}
\end{align}
Here $Q$ is the Carter constant famously allowing for the separability
of the equation of motion $\frac{\partial S}{\partial Q}=0$. Namely, $p_x$ and $p_\rho$ are functions of $x$ and $\rho$ respectively. From either of the above equations, the Carter constant can be written as a quadratic form in the momenta, preserved under the Hamiltonian flow,
\begin{equation}
Q=Q^{\mu\nu}p_{\mu}p_{\nu},\,\,p_{t}=\omega\,,\,\dot{Q}=0\,,
\end{equation}
where $Q^{\mu\nu}$ is a Killing tensor. This is the quintessential hidden symmetry of type D backgrounds, of which the Kerr metric is the most interesting case.  Now, even though
decoupled and classically integrable, the radial and angular integrals in (\ref{eq:spinxq})
are elliptic functions underlying the chaotic motion of an orbit in a rotating background. In order to obtain
an explicit result its numerical evaluation is often required. However,
a simplification appears at the SD point as the Killing tensor becomes
reducible and the symmetry becomes explicit. To be more precise, one finds
\begin{equation}
Q^{\mu\nu}\to K^{\mu\nu}+aA_{3}^{\mu\nu}\,,\,\,\det(K^{\mu\nu})=0\,,\,\,\,\textrm{as}\,\,N\to M\,.
\end{equation}
Here $A_{3}^{\mu\nu}$ is the $z$-component of the famous Laplace-Runge-Lenz
triplet.\footnote{See \cite{Guevara:2023wlr} for an explicit expression. Away from the SD point, the averaged
time evolution of $A_{3}=A_{3}^{\mu\nu}p_{\mu}p_{\nu}$ is nothing
but the precession $\Delta\Phi$ found in the previous section.} On the other hand, one finds that $K^{\mu\nu}$ is a symmetric rank-3 Killing tensor, therefore
it can be written as
\begin{equation}
    K=\sum_{i=1}^{3}\xi_{i}\xi_{i}\,.
\end{equation}
These are
nothing but a triplet of Killing vectors $\xi_{i}^{\mu}$ generating the emergent
$SO(3)$ symmetry for the SD Kerr metric. Explicitly, we denote
\begin{equation}
J^{2}:=K^{\mu\nu}p_{\mu}p_{\nu}=p_{\phi}^{2}+\frac{(2M(ax-\rho)\omega+(a-\rho x)p_{\phi})^{2}}{\Delta(\rho)(x^{2}-1)}+\frac{\Delta(\rho)(x^{2}-1)(ap_{\rho}+p_{x})^{2}}{(\rho+ax)^{2}}\label{eq:cas}
\end{equation}
(here $\Delta(\rho)=\rho^{2}-a^{2}$, namely we have shifted $\rho\to\rho+M$
with respect to (\ref{eq:sd1xa})). In analogy to equation \eqref{eq:tfind}, setting
our frame as $p_{\phi}=J$ implies the last two terms vanish separately,
i.e.
\begin{align}
\frac{\rho x-a}{ax-\rho} & =2M\omega/p_{\phi}\,,\nonumber \\
p_{x} & =-ap_{\rho}\,.\label{eq:fna2}
\end{align}
The first condition implies that on the trajectory we can algebraically solve $x=x(\rho)$.
The second condition implies that the radial and angular problems
can now be combined
\begin{equation}
\int p_{x}dx+\int p_{\rho}d\rho=\int p_{\rho}(\rho)(d\rho-adx)\,, \label{eq:xfor}
\end{equation}
The value of the Carter constant $Q$ is further solved from (\ref{eq:pxprho})-(\ref{eq:fna2}), which gives
\begin{equation}
Q=p_{\phi}^{2}+4M^{2}\frac{a\omega}{p_{\phi}}(m^{2}-2\omega^{2})\,.\label{eq:asz1}
\end{equation}
Motivated by the form \eqref{eq:xfor}, let us now introduce $dr=d\rho - adx$, namely
\begin{equation}
r:=\rho-ax(\rho)=\frac{a^{2}-\rho^{2}}{a\frac{2M\omega}{p_{\phi}}-\rho}\,.\label{eq:ncch}
\end{equation}
Further using (\ref{eq:asz1}) and (\ref{eq:pxprho})
(for $N=M$) we find, after some algebra,

\begin{equation}
p_{\rho}=\frac{\sqrt{\mathcal{P}(\rho+M)}}{\rho^{2}-a^{2}}=\frac{\sqrt{p_{\phi}^{2}+r(r+4M)\omega^{2}-r(r+2M)m^{2}}}{r}
\end{equation}
which can be shown to agree with the static radial action \eqref{eq:sam1}. This shows that
we have reduced the Kerr problem to the static SDBH case. We conclude
that the rotating metric becomes integrable at the SD point. In hindsight, the coordinate change \eqref{eq:ncch} can be derived from the large diffeomorphism found in \cite{Crawley:2021ivb}, which removes the spin dependence at the SD point. This can be seen as an off-shell version of the famous Newman-Janis shift. In the following, we mostly specialize in the case $a=0$ and comment on how to produce the $a\neq 0$ results from this method.

\section{Quantum amplitude from WKB formulation}\label{sec:wkb}

In the next two sections we will argue that the classical point particle trajectories we have derived are enough to determine an exact scattering amplitude on this background. The relation
will be given by a semiclassical formula very closely related to
the WKB expansion. For massive modes the amplitude is extracted from the asymptotic expansion of the Euclidean wavefunction. Its precise analytic continuation as well as the massless modes are discussed in section \ref{sec:contd}.

As a warm up, we will consider in this section
the quantum amplitude by using the WKB ansatz
\begin{align}
\psi & =Ae^{\frac{i}{\hbar}S}\nonumber \\
 & =e^{i\omega t}A_{\theta}e^{\frac{i}{\hbar}S_{\theta,\phi}}A_{r}e^{\frac{i}{\hbar}S_{r}}\,,
\end{align}
and expand the wave equation
\begin{equation}
\partial^{2}\psi=\frac{m^{2}}{\hbar^{2}}\psi \,,\label{eq:kge}
\end{equation}
in powers of $\hbar$. The leading order is quadratic in the on-shell action $S$, namely the phase, and it is given by
the HJ equations \eqref{eq:hjeq}. For general $M$ and $N$, the $\mathcal{O}(\hbar)$ corrections
yield the amplitudes \footnote{The $\sim\hbar^{2}$ equation will not be considered here. For completeness, it reads $\sin\theta(\partial_{\theta}\log A)^{2}+\partial_{\theta}(\sin\theta\partial_{\theta}\log A)=0$.}
\begin{align}
 A_{\theta} & =\frac{1}{\sqrt{\sin\theta p_{\theta}}}\,,\nonumber \\
A_{r} & =\frac{1}{\sqrt{p_{r}}\Delta^{\frac{1}{4}}} \sim\frac{1}{r\sqrt{p}} \,.
\end{align}
For the usual angular problem in the absence of monopole charges (namely $p_{\theta}=J$), Ford and Wheeler 
noted long ago that the partial wave expansion is directly connected
to a WKB approximation \cite{Ford1959}. Based on further seminal work by Schwinger
et al \cite{Schwinger1976}, recently \cite{Kol:2021jjc} extended the analysis to monopole
scattering by considering the monopole harmonics, namely exact solutions of the angular equation in \eqref{eq:kge}. The function $A_{\theta}e^{iS_{\theta,\phi}}$
agrees with the result from asymptotics of the monopole harmonics.\footnote{The difference here is that we did not need the explicit formula for
monopole harmonics. This could be relevant in cases where the angular
solution is not elementary such as Kerr. } 

In this section we will define the momentum amplitude guided by the
Ford-Wheeler approach and show that it is almost exact. The incoming
plane wave impinges in the $z$ direction, leading to the usual expansion
\begin{equation}
f(\theta)=\frac{\psi(r\to\infty)}{(2ipA_{r}e^{i\omega t-ipr+MP\log...})}=\frac{1}{2ip}\sum_{J}(2J+1)A_{\theta}e^{iS_{\theta,\phi}}e^{i\bar{S}_{r\to\infty}}\,.\label{eq:pwam}
\end{equation}
In $\bar{S}_{r\to\infty}$
we keep the $J$ dependent terms of $2S_{r\to\infty}$, since the remainder is an overall phase. Now recall that total and orbital angular momenta are related via
\begin{equation}
    J^2=L^2+q^2\,.
\end{equation}
For the position scattering amplitude it is convenient to introduce the impact parameter $L=\omega b$.
The eikonal/semiclassical
limit is
\begin{equation}
    \omega=\frac{E}{\hbar}\to\infty\,,\quad q/L \,\,\,\textrm{fixed.}
\end{equation}
On the other hand $b=L/\omega$ has resolution $1/\omega$ and thus can be
taken as continuum. The eikonal approximation of the partial wave amplitude \eqref{eq:pwam} then becomes

\begin{equation}
f_{\textrm{approx}}(\theta)=\frac{1}{ip}\int\frac{JdJ}{\sqrt{\sin\theta p_{\theta}}}e^{iS_{\theta}}e^{i(\bar{S}_{r\to\infty}(J))}\,.\label{eq:pwqe}
\end{equation}

\subsection{Cross section}

Let us first show that this expression recovers a good approximation
of the cross section for a general angular potential and radial potentials
$p_{\theta},p_{r}$, namely for generic mass and NUT charges. The saddle of \eqref{eq:pwqe} is
\begin{equation}
f_{\textrm{approx}}(\theta)\sim\frac{1}{ip}\frac{J}{\sqrt{\pi\sin\theta p_{\theta}\left(\frac{\partial^{2}\bar{S}_{r}}{\partial J^{2}}+\frac{\partial^{2}S_{\theta}}{\partial J^{2}}\right)_{\theta_{J}}}}e^{iS_{\theta_{J}}}e^{i\bar{S}_{r\to\infty}(J)}\,.
\end{equation}
Recalling the definition $\beta(\theta,J)=-\frac{\partial S_{\theta}}{\partial J}$ from the previous section,
the saddle is evaluated at a classical scattering angle $\theta=\theta_{J}$ satisfying
\begin{equation}
\frac{\partial\bar{S}_{r}}{\partial J}=\beta(\theta_{J},J)\label{eq:sdpar}
\end{equation}
Regarding both sides as functions of $J$ 
\begin{equation}
\frac{\partial^{2}\bar{S}_{r}}{\partial J^{2}}=\frac{\partial\beta(\theta,J)}{\partial\theta}\frac{d\theta_{J}}{dJ}-\frac{\partial^{2}S_{\theta}}{\partial^{2}J}\qquad
\text{at} \qquad \theta=\theta_{J}.
\end{equation}
Plugging this into the saddle we get 
\begin{equation}
f_{\textrm{approx}}(\theta)\sim\frac{1}{ip\sqrt{\pi}}\left(\frac{p_{\theta}}{J}\frac{\partial\beta}{\partial\theta}\right)^{-1/2}\sqrt{\frac{JdJ}{\sin\theta d\theta}}e^{iS_{\theta_{J}}}e^{i\bar{S}_{r\to\infty}(J)}
\end{equation}
We need to show that the term in parenthesis equals unity. 
For this, note that a generic angular potential takes the form $p_{\theta}^{2}=J^{2}+\ldots$, where the corrections can involve the monopole charges. Using again $\beta(\theta,J)=-\frac{\partial S_{\theta}}{\partial J}$ we find
\begin{equation}
\frac{p_{\theta}}{J}\frac{\partial\beta}{\partial\theta}=\frac{p_{\theta}}{J}\frac{\partial}{\partial\theta}\frac{\partial}{\partial J}S_{\theta}=\frac{p_{\theta}}{J}\frac{\partial}{\partial J}p_{\theta}=\frac{\partial p_{\theta}^{2}}{\partial J^{2}}=1\,,
\end{equation}
as desired. 
Now, since $JdJ=LdL=p^{2}bdb$ this shows that the amplitude is well approximated by
\begin{equation}
f_{\textrm{approx}}(\theta)\sim\frac{1}{i\sqrt{\pi}}\underbrace{\sqrt{\frac{bdb}{\sin\theta d\theta}}}_{d\sigma/d\Omega}e^{iS_{\theta_{J}}}e^{i\bar{S}_{r\to\infty}(J)}\label{eq:sarp}
\end{equation}
In other words, $|f(\theta)|^2$ coincides with the classical cross-section. This generalizes the standard result to the case where a monopole/NUT charge enters the angular problem.

\subsection{Phase}

Having verified that the cross section is recovered we now move on to
compute the phase in the eikonal limit. We will specialize to 
the SD point since the calculation can be completely controlled. Indeed, we anticipate the exact amplitude will be derived in the next two sections from different methods, but we find it useful to quote this part of the result here. It turns out that by solving the Klein-Gordon equation in suitable coordinates we can obtain the exact phase shift:
\begin{equation}
\delta=\mathcal{M}p\log\sin^{2}\theta/2+i\log\frac{\Gamma(1+q+i\mathcal{M}p)}{\Gamma(1+q-i\mathcal{M}p)}+\arctan\frac{\mathcal{M}p}{q}\label{eq:2ts}
\end{equation}
Clearly, we expect the first two terms to survive the eikonal limit
whereas the last to be invisible since it is $\mathcal{O}(1)$.

As $r\to \infty$, the saddle point \eqref{eq:sdpar} is
\begin{equation}
\arctan(\frac{J\cot\theta/2}{p_{\theta}})=\arctan(\frac{J}{\mathcal{M}p})\,,\label{eq:sadle}
\end{equation}
which is solved by the scattering angle
\begin{equation}
\sin^{2}\theta/2=\frac{\mathcal{M}^{2}p^{2}+q^{2}}{\mathcal{M}^{2}p^{2}+J^{2}}\,.
\end{equation}
Plugging this back into \eqref{eq:sarp}, using the form of the actions \eqref{eq:rct} and \eqref{eq:angact}, we get the following phase
\begin{align}
i(-\bar{S}_{r}+S_{\theta}) & =2i\mathcal{M}p+i\mathcal{M}p\log(\mathcal{M}^{2}p^{2}+J^{2})-2iq\arctan\left(\frac{q\cot\theta/2}{p_{\theta}}\right)\nonumber \\
 & =2i\mathcal{M}p-i\mathcal{M}p\log\sin^{2}\theta/2+(i\mathcal{M}p+q)\log(i\mathcal{M}p+q)-(q-i\mathcal{M}p)\log(q-i\mathcal{M}p) \label{eq:srwac}
\end{align}
We will use the Stirling approximation $z\log z\sim\log\Gamma(z+1)-z$
for large $z$ to write

\begin{equation}
i(-\bar{S}_{r}+S_{\theta})=\log\frac{\Gamma(1+q+i\mathcal{M}p)}{\Gamma(1+q-i\mathcal{M}p)}-i\mathcal{M}p\log\sin^{2}\theta/2
\end{equation}
This recovers the first two terms of (\ref{eq:2ts}) as expected. The eikonal phase shift dominates in the high-energy limit and agrees with the well known Newtonian scattering phase as $q\to 0$ (see e.g. \cite{Bautista:2021wfy}). We emphasize however that this is a correction to all orders in $GM\omega$.

\section{The Post-Minkowskian formula}\label{sec:pmf}

We saw in the previous sections that the the self-dual black hole is a candidate for a solvable system and generalizes the integrability of the Kepler problem. More precisely, the observables can be computed to any order in $GM/b$, the so-called Post-Minkowskian amplitude.  

Part of the simplicity stems from a very simple functional dependence of the radial problem controlled by $p_r$. After all, a tower of PM corrections of type $GM/r$ are absorbed in the shift $r\to r+M$ in $p_r$ (see \eqref{eq:rct}) after which the effective potential simply becomes
\begin{equation}
V(r)=\frac{2\mathcal{M}p^{2}}{r}=\frac{2M(\omega^{2}-p^{2})}{r}\label{eq:sda1}
\end{equation}
As a function of $r$ this is indeed the 3d Fourier transform of the first Post-Minkowskian amplitude,
namely (relativistic) Newtonian gravity or simply the linearized Schwarzschild potential.\footnote{Interestingly, it does not match the 1PM amplitude obtained in self
dual gravity, a theory with only positive helicity gravitons. To be precise, the 1PM amplitude in gravity, continued
to Euclidean signature, is

\begin{align}
\mathcal{M} & =\frac{M^{2}}{k^{2}}\left((\omega+ip)^{2}+(\omega-ip)^{2}\right)\label{eq:amp2}\\
 & =-\frac{Mp}{k^{2}}\left((\mathcal{M}p-iq)+(\mathcal{M}p+iq)\right)
\end{align}
i.e. the Fourier transform of (\ref{eq:sda1}). Here each of the two
terms is one of the on-shell graviton exchanges. Hence the SD amplitude
only recovers the first term. As it will be clear in the next section,
the missmatch is expected since the procedure to obtain the scattering
angle in this note (from (\ref{eq:sda1})) is a modified version of
the map used on the amplitude (\ref{eq:amp2}) (see \cite{Emond:2020lwi}
for the latter).} In this picture, corrections from the NUT charge play two roles:
\begin{itemize}
\item Modifying the angular problem and
\item cancelling higher Post-Minkowskian
effects in (\ref{eq:asdw}) to make the potential integrable to all
orders in $G$. 
\end{itemize}
It turns out that once we take into account the modified
angular problem we can obtain a formula for the scattering amplitude
to all orders in $G$, based solely in (\ref{eq:sda1}).

We will provide a simple argument and leave detailed derivation for
later work. The idea is to consider an equivalent Schrodinger equation 
\begin{equation}
H_{0}+V=\left[(\partial_{i}+qA_{i})^{2}-\frac{q^{2}}{r^{2}}\right]+V(r)=p^{2}\,.\label{eq:scwa}
\end{equation}
This can be regarded as the electromagnetic dyon equation if the potential is chosen
as 
\begin{equation}
A_{i}dx^{i}=(\zeta-\cos\theta)d\phi \,,\label{eq:dp}
\end{equation}
and $V$ is given by (\ref{eq:sda1}). The radial and angular components
of this problem precisely match the SD black hole if we choose $q=-2iN\partial_{t}$.\footnote{Related observations recently appeared in \cite{Adamo:2023fbj,Guevara:2023wlr}.
In the later reference we argued that the \textit{radial}
problem can be mapped to the Coulomb problem as it is evident from
(\ref{eq:scwa}). }

The free Hamiltonian $H_{0}$ is simply the well known monopole hamiltonian
plus a $\frac{q^{2}}{r^{2}}\sim G^{2}$ term which we ignore in the
Post-Minkowskian limit. In the next section we argue that the exact
$H_{0}$ operator is indeed a free relativistic Hamiltonian. Here
instead we want to approximate a solution accurate to linear order
in $G$ that is propagating in the $z$ direction. We will take

\begin{align}
\psi_{p} & =\exp i(pz+qA_{\phi}\phi+\frac{1}{2p}\int^{z}V(b,\phi,z')dz')\label{eq:asd2}\\
\nonumber 
\end{align}
where $b^{2}+z^{2}=r^{2}$. We will approximate $qA_{\phi}\phi\approx2q\phi$
for large values of $z$. Indeed, at large $z\approx r$ we can neglect
the angular components and verify that the plane wave matches the
radial action (corresponding to a spherical wave)
\begin{align}
S_{r} & =\int\sqrt{p^{2}-\frac{J^{2}}{r^{2}}+V(r)}dr\approx pr+\frac{1}{2p}\int V(r)dr\nonumber \\
 & \approx pz+\frac{1}{2p}\int V(b,z)dz
\end{align}
Classically, this action represents almost straight trajectories in
the $z$ direction with very little deflection\footnote{A detailed derivation of (\ref{eq:asd2}) for the case $A=0$ can be found in \cite{Glauber1959}. }. Equivalently, the action is linear in $G$. As it is well known,
for the Coulomb problem the integral needs to be regulated

\begin{equation}
\chi(b)=\frac{1}{2p}\int_{-1/\epsilon}^{1/\epsilon}V(b,z)dz=\int_{-1/\epsilon}^{1/\epsilon}\frac{\mathcal{M}p}{\sqrt{b^{2}+z^{2}}}dz=2\mathcal{M}p\log\frac{2\epsilon}{pb}+\mathcal{O}(\epsilon^{2})\label{eq:exin}
\end{equation}
As expected, this is the 1PM value of the radial action if we identify
$J\propto b$. It can also be shown to agree with \eqref{eq:srwac} in terms of the scattering angle. Our new eikonal formula is obtained by plugging this
approximation into the well-known Lippman-Schwinger equation that provides the scattering amplitude,
\begin{align}
f(p,p') & =\frac{i}{4\pi}\int d^{2}bdze^{-ip'\cdot x}V(b,\phi,z)\psi_{p}(b,\phi,z)\nonumber \\
 & =\frac{i}{4\pi}\int d^{2}bdze^{ik\cdot x}V(b,\phi,z)e^{i2q\phi}e^{\frac{i}{2p}\int_{-\infty}^{z}V(b,\phi,z')dz'}\nonumber \\
 & =\frac{p}{2\pi}\int d^{2}be^{ik\cdot x}e^{i2q\phi}e^{\frac{i}{2p}\int_{-\infty}^{\infty}V(b,\phi,z')dz'}
\end{align}
where $k=p-p'$ is a momentum transfer. Since $x=b+z\hat{p}$ we can
take $k\cdot x\approx k\cdot b$. We then obtain the eikonal formula as quoted in the introduction:
\begin{equation}
f(k)=\frac{p}{2\pi}\int d^{2}be^{ikb+i2q\phi}e^{i\chi(b)}\,.
\end{equation}
This generalizes the usual eikonal formula to the case $q\neq0$. The new phase term $e^{2iq\phi}$ stems from the corrected angular problem, therefore we expect the formula to hold for general NUT charges beyond the SD point. This introduces a full tower of PM corrections in $q=2N\omega$. For our discussion we will apply it to the SDBH since the radial factor $e^{i\chi(b)}$ is also known to all orders.

Inserting the 1PM on-shell action from \eqref{eq:exin},
where we discard the `Newton' phase $\Phi_{N}=2\mathcal{M}p\log2\epsilon$,
we get
\begin{align}
\frac{2\pi f(k)}{p} & =\int d^{2}be^{ikb+i2q\phi}e^{i\chi(b)}\nonumber \\
 & =\int bdbd\phi e^{i2pb\sin\theta/2\cos\phi}e^{i2q\phi}e^{i\chi(b)}\nonumber \\
 & =2\pi e^{i\pi q}\int bdbJ_{2q}(2pb\sin\theta/2)e^{i\chi(b)}\nonumber \\
 & =\frac{2\pi e^{i\pi q}}{4p^{2}\sin^{2}\theta/2}e^{i\mathcal{M}p\log4\sin^{2}\theta/2}\int bdbJ_{2q}(b)b^{i2\mathcal{M}p}.\label{eq:derca}
\end{align}
Using the following formula for integral of the Bessel function
\begin{equation}
\int bdbJ_{2q}(b)b^{\alpha}=\frac{2^{1+\alpha}\Gamma(1+\alpha/2+q)}{\Gamma(q-\alpha/2)}\,,\label{eq:soma}
\end{equation}
we obtain
\begin{align}
f(k) & =\frac{e^{i\pi q}}{2p\sin^{2}\theta/2}e^{i\mathcal{M}p\log\sin^{2}\theta/2}(q-i\mathcal{M}p)\frac{\Gamma(1+q+i\mathcal{M}p)}{\Gamma(1+q-i\mathcal{M}p)}\,.\label{eq:12x}
\end{align}
The phase and modulus of this amplitude fully agree with the description
of the previous section. On the other hand, we will recover this expression
by solving the wave equation exactly in the next section.

Finally, we would like to comment about the meaning of the impact parameter $b$ which is sometimes ambiguous. The saddle approximation of the second line of (\ref{eq:derca}) is 
\begin{align}
\cos\phi & =\frac{\sqrt{p^{2}b^{2}\sin^{2}\theta/2-q^{2}}}{pb\sin\theta/2}\,,\nonumber \\
\sin\theta/2\sqrt{p^{2}b^{2}-q^{2}\csc^{2}\theta/2} & =-\mathcal{M}p\,.
\end{align}
The latter agrees with (\ref{eq:sadle}) at small angles and hence
we can identify $L=pb$. In the 2d $b$-plane, these can be written
as
\begin{equation}
k=\frac{2q}{b}\hat{p}+\frac{2\mathcal{M}p}{b}\hat{b}\,,
\end{equation}
where $\hat{p}\cdot\hat{b}=0$. Indeed, for small angles we obtain
\begin{equation}
\theta=2\frac{\sqrt{\mathcal{M}^{2}+\frac{q^{2}}{p^{2}}}}{b}=\frac{2M}{b}(1+\omega^{2}/p^{2})\,.
\end{equation}
But our formula is exact, and thus $b$ in general does not match
the classical impact parameter at infinity. 

\section{The exact amplitude from Complex structure}\label{sec:comps}

We will now show that the amplitude can be computed directly from the
wave equation thanks to the enriched separability on this background. This results from a particular set of complex coordinates which are familiar from the phase space of the hydrogen atom\footnote{They were used long ago in solving the wave equation of the BPS monopole \cite{Gibbons:1986df}, a particular case $(M=-1,a=0)$ of the background discussed in this section. Our analysis in this section is heavily inspired by that seminal work but we do not rely on the existence of the LRL triplet.}, yet in the SDBH they are genuine spacetime coordinates.  

It is illustrative to slightly generalize our discussion to include spin $a\neq0$.
Consider again the Euclidean Kerr metric with NUT charge \eqref{eq:sd1xa}.
%
%\begin{equation}
%ds^{2}=\Sigma(\frac{dr^{2}}{\Delta}-%d\theta^{2})+\frac{\sin^{2}\theta}{\Sigma}(adt-\rho %d\phi)+\frac{\Delta}{\Sigma}(dt+Ad\phi)^{2}\label{eq:sdf2}
%\end{equation}
%The functions $\Sigma,\Delta,A,\rho$ can be found in {[}/2311.07933{]}.
In\cite{Guevara:2023wlr} we argued that this metric admits a very simple coordinate
set at the SD point which absorbs the rotation parameter $a$. It
is given by
\begin{align}
z_{+} & =\sqrt{r-M+a}\cos\theta/2e^{\frac{it}{4M}}e^{-i\frac{\phi}{2}}\nonumber \\
z_{-} & =\sqrt{r-M-a}\sin\theta/2e^{\frac{it}{4M}}e^{i\frac{\phi}{2}}\label{eq:cazz}
\end{align}
In the case $a=0$ these coordinates are closely related to the hyper-Kähler structure
of the Taub-NUT solution \cite{Gibbons1988}. The $\pm$ index represents an spinor index
under the `hidden' rotation group of the metric \eqref{eq:sd1xa}, which is $SU(2)$ with $J_{z}=\partial_{\phi}$. Its Casimir was given in \eqref{eq:cas}.

Remarkably, the Klein-Gordon equation in this background can be written
very simply. After working out the coordinate change \eqref{eq:cazz} in the metric \eqref{eq:sd1xa}, it turns into

\begin{equation}
\partial_{+}\bar{\partial}_{+}\psi+\partial_{-}\bar{\partial}_{-}\psi+p^{2}(z_{+}\bar{z}_{+}+z_{-}\bar{z}_{-})\psi=-2\mathcal{M}p^{2}\psi\label{eq:dxa}
\end{equation}
As anticipated, the spin parameter is simply irrelevant. To understand this structure, let us reconsider the Taub-NUT case, namely \eqref{eq:tnsx}. Recall that we will shift $r\to r+M$ and set the
gauge as $\zeta=0$. The coordinates (\ref{eq:cazz}), when specialized to $a=0$, have the property
that the Hamiltonian in (\ref{eq:scwa}) becomes the free Laplacian
\begin{equation}
rH_{0}=r(\partial_{i}+A_{i})^{2}-\frac{q^{2}}{r}=\partial_{+}\bar{\partial}_{+}+\partial_{-}\bar{\partial}_{-}
\end{equation}
Thus (\ref{eq:dxa}) precisely agrees with the Schrodinger problem
(\ref{eq:scwa}) (with $V(r)$ given by \eqref{eq:sda1} and $r=z_{+}\bar{z}_{+}+z_{-}\bar{z}_{-}$ in that
case).

The rectangular structure in \eqref{eq:dxa} immediately makes it obvious that there
exists separable solutions transforming as $SU(2)$ representations.
A simple choice is

\begin{align}
\psi & =\left(\frac{z_{+}z_{-}}{\bar{z}_{+}\bar{z}_{-}}\right)^{q/2}f_{+}(|z_{+}|^{2})f_{-}(|z_{-}|^{2})
\end{align}
Recalling that $q=2M\omega$, the prefactor $\sim e^{i\omega t}$
simply accounts for the energy, while the modulus $\xi_{\pm}=|z_{\pm}|^{2}$
is so that the solution does not depend on the angle $\phi$, matching
the analysis of the previous section. To obtain the equation in a
different gauge $\zeta\neq0$, we can simply transform $t\to t+2N\zeta\phi$ in
this solution rather than transforming the equation, leading to

\begin{equation}
\psi_{\zeta}=\left(\frac{z_{+}z_{-}}{\bar{z}_{+}\bar{z}_{-}}\right)^{q/2}\left(\frac{\bar{z}_{+}z_{-}}{z_{+}\bar{z}_{-}}\right)^{\zeta q/2}f_{+}(|z_{+}|^{2})f_{-}(|z_{-}|^{2})\,.
\end{equation}
The coordinates $\xi_{\pm}$ are precisely the parabolic coordinate
system familiar from the treatment of the hydrogen atom. We obtain
the following decoupled system
\begin{equation}
(\xi_{\pm}f'_{\pm})'+\left[(\mathcal{M}+\xi_{\pm})p^{2}-\frac{q^{2}}{4\xi_{\pm}}(\zeta\pm1)^{2}\pm\lambda\right]f_{\pm}=0
\end{equation}
As it is known for parabolic coordinates, the eigenvalue $\lambda$
is related to the Casimir of a Laplace-Runge-Lenz vector discussed in section \ref{sec:kerr}. 

By choosing the gauge $\zeta=-1$ as in the previous section, one of
these equations admits a plane wave solution suitable to describe
scattering

\begin{equation}
f_{+}=e^{ip\xi_{+}} ,\qquad \textrm{with} \qquad\lambda=-ip-\mathcal{M}p^{2}\,.
\end{equation}
The second equation becomes
\begin{equation}
(\xi_{-}f'_{-})'+\xi_{-}p^{2}f_{-}+\left[2\mathcal{M}p^{2}-\frac{q^{2}}{\xi_{\pm}}+ip\right]f_{-}=0\,,
\end{equation}
which admits the following scattering solution in terms of the generalized
Laguerre polynomials:
\begin{align}
f_{-}^{\textrm{scat}} & =e^{-ip\xi_{-}}(-2ip\xi_{-})^{q}L_{-i\mathcal{M}p-q}^{2q}(2ip\xi_{-})\nonumber \\
 & \sim e^{-ip\xi_{-}-i\mathcal{M}p\log2p\xi_{-}}+(-1)^{i\mathcal{M}p}\frac{1}{(2p\xi_{-})}\frac{\Gamma(1+q-i\mathcal{M}p)}{\Gamma(q+i\mathcal{M}p)}e^{ip\xi_{-}+i\mathcal{M}p\log2p\xi_{-}}\,.
\end{align}
After adding the factor $f_{+}=e^{ip\xi_{+}}$, the first term leads
to a plane wave propagating in the $z\sim r\cos\theta$ direction
and the second is the scattering amplitude. 
\begin{align}
f_{-}f_{+} & \sim e^{ipr\cos\theta}e^{-i\mathcal{M}p\log2pr}e^{-i\mathcal{M}p\log\sin^{2}\theta/2}\nonumber \\
 & +(-1)^{i\mathcal{M}p}\frac{r}{2p\xi_{-}}e^{ipa(\cos\theta-1)}e^{i\mathcal{M}p\log\sin^{2}\theta/2}\frac{\Gamma(1+q-i\mathcal{M}p)}{\Gamma(q+i\mathcal{M}p)}\frac{e^{ipr+i\mathcal{M}p\log2pr}}{r}
\end{align}
Recalling that the spin is aligned with the $z$ direction\footnote{This is the polar scattering considered in \cite{Bautista:2021wfy}. We leave
the analysis of general spin orientation, also considered there, for
future work. }, we can write $e^{ipa(\cos\theta-1)}=e^{ia\cdot k}$ which leads to
the full amplitude
\begin{equation}
\mathcal{M}=\frac{e^{ia\cdot k}}{2p\sin^{2}\theta/2}e^{i\mathcal{M}p\log\sin^{2}\theta/2}(q+i\mathcal{M}p)\frac{\Gamma(1+q-i\mathcal{M}p)}{\Gamma(1+q+i\mathcal{M}p)}\label{eq:someamp}
\end{equation}
in agreement (\ref{eq:12x}) with for $a=0$. The spinning case implements
the Newman-Janis shift by means of the factor $e^{ia\cdot k}$, as
outlined in \cite{Arkani-Hamed:2019ymq} and \cite{Guevara:2018wpp}. In particular the tree-level version of the
amplitude
\begin{align}
\mathcal{M_{\textrm{tree}}} & =\frac{2ip}{k^{2}}e^{ia\cdot k}(\mathcal{M}p-iq)\label{eq:amp2-1}
\end{align}
matches previous results \cite{Huang:2019cja} when analytically
continued to Euclidean signature. We note the absence of the 1-loop
correction $\sqrt{k^{2}}\propto\sin\theta/2$ \cite{Guevara:2017csg} consistent with
the expectation that the 1-loop amplitudes for the two-body problem
vanish in self-dual gravity, see discussion.

\section{Kleinian scattering and massless amplitude}\label{sec:contd}

Working in Euclidean signature has allowed to derive a striking connection
between the self-dual black hole and a non-relativistic Schrodinger
problem (\ref{eq:scwa}). However, this analysis lacks key scattering
features such as radiative modes at null infinity. A novel description
of self-dual black holes and its radiative modes has been given in
\cite{Crawley:2023brz} for Klein signature, also known as the split $(2,2)$ signature. This signature is accesible
to us via the following analytic continuation
\begin{align}
\theta & \to i\theta , \nonumber \\
J & \to iJ, \\
p & \to ip .\nonumber 
\end{align}
The first and second transformations simply unwind the coordinate
$\theta\geq0$ which is now non-compact. We see that in the new metric
\begin{equation}
ds^{2}=\frac{\Delta(r)}{r^{2}-N^{2}}\left(dt+2N(\zeta-\cosh\theta)d\phi\right)^{2}+(r^{2}-N^{2})\left(dr_{*}^{2}-d\theta^{2}-\sinh^{2}\theta d\phi^{2}\right)
\end{equation}
the 3d spatial part is now Lorentzian and corresponds to $AdS_{3}$.
The last coordinate transformation allows us to consider massless
modes located at $p=\pm\omega$. Indeed, consider the new momentum
relations
\begin{align}
p_{\theta}^{2}+\frac{(J_{z}+q\cosh\theta)^{2}}{\sinh^{2}\theta} & =J^{2}+q^{2}\,\,\,\label{eq:asdw-1}\\
-p_{r}^{2} & +J^{2}=p^{2}r^{2}+2p^{2}r\mathcal{M}\nonumber 
\end{align}
which shows that $p^{2}$ is not sign-definite. Null geodesics are
obtained at $p=\pm\omega$.

\subsection{Lyapunov exponents}
Besides the precession calculation of section \ref{sec:prc}, a second test of classical and quantum integrability emerges in the massless case from the consideration of the Lyapunov exponents of homoclinic trajectories. As expected, these unstable null trajectories dissapear at the self-dual point, reflected in the Lyapunov exponent diverging. For a null geodesic, the observable significance of this quantity is no other than a characterization of the black hole photon ring \cite{Lupsasca:2018tpp,Yang:2012he}. We will develop the relation to scattering amplitudes, in particular quasi-normal modes, for these NUT-charged black holes in \cite{upcoming:2025}, here we will instead address the classical orbits.

The definition of a Lyapunov exponent of a trajectory depends on its affine parametrization. Usually this is taken to be the coordinate time or the angular period. For non-static solutions this involves averaging over angular integrals. We will simplify part of the analysis by choosing a suitable affine parameter
\begin{equation}
    d\lambda=\partial_J p_r dr_{*} = - \partial_J p_{\theta}\,d\theta
\end{equation}
(the equality follows from the HJ equation). From here it follows that
\begin{align}
    \frac{dr}{d\lambda}=&\frac{\sqrt{\Delta(r)}}{\partial_J p_r} 
\end{align}
We now read off the unstable radius. Specifically, let $p_r=\sqrt{\mathcal{V}(r)}$ so that
\begin{equation}
    \frac{dr}{d\lambda}=\frac{2\sqrt{\Delta(r)\mathcal{V}(r)}}{\partial_J\mathcal{V}(r)}  \label{eq:twrad}
\end{equation}
and look for the maximum of the potential $\mathcal{V}(r)$:
\begin{equation}
    \mathcal{V}(r_0)=\mathcal{V}'(r_0)  =0
\end{equation}
The solution can be understood as a phase space point $(r_o,\omega_0)$ (the latter dependence is implicit in the above equation). To write it, it is convenient to introduce
\begin{align}
    \kappa_{\pm}:=&(M\pm N)^{1/3}\,,\\
    \mathcal{K}:=&k_+^2+k_+k_-+k_-^2\,.
\end{align}
The former are analog to the Hawking temperature for the Kleinian solutions \cite{upcoming:2025}. Then
\begin{align}
    r_0 =& \frac{1}{2}(k_+ + k_-)\mathcal{K}\\
    \frac{J}{\omega_0} =& \sqrt{3k_+ k_-}\mathcal{K}
\end{align}
At $r=r_0+\delta r$ the EOM \eqref{eq:twrad} then reads
\begin{equation}
    \frac{1}{\delta r}\frac{d\delta r}{d\lambda}= \sqrt{1+\frac{2(k_++k_-)\mathcal{K}\delta r+\delta r^2}{3k_+k_-\mathcal{K}^2}}
\end{equation}
For finite $k_\pm$ and $\delta r\ll k_\pm^3$ we see that the trajectory is integrated as
\begin{equation}
    \delta r \sim e^{\lambda} \delta r_{\lambda=0}
\end{equation}
signaling chaotic trajectories around the unstable maximum. However, as $k_- \to 0$ or $k_+ \to 0$ this expansion breaks down: We find $\delta r$ grows only polynomially as $\lambda\to \infty$.

\subsection{Lorentzian Amplitude}

Similarly, a continuation to Lorentzian signature from Euclidean signature
proceeds via a standard Wick rotation
\begin{align}
t & \to -it , \nonumber \\
\omega & \to i\omega, \\
N & \to -iN , \nonumber \\
a & \to -ia .
\end{align}
The latter reflecting $N$ and $a$ being `pseudoscalars'. The combinations $q=2N\omega$ and $a\omega$ remain invariant (up to a sign flip).

The SD point is at $M=iN$. This
conditions prevents the SD metric from being real, but we expect
the S-matrix to be defined via analytic continuation anyway. Further analysis is required to address its causality.

We can now discuss the massless modes of the SD background. Taking $p^{2}=\omega^{2}$ in the continuation of the amplitude \eqref{eq:someamp} we find

\begin{equation}
\mathcal{\mathcal{A}}=\frac{ie^{ia\cdot k}}{2p\sin^{2}\theta/2}e^{i2Mp\log\sin^{2}\theta/2}(2M(\omega+p))\frac{\Gamma(1+i2M(\omega-p))}{\Gamma(1+i2M(\omega+p))}.
\end{equation}
This is expected to resum the Teukolsky amplitudes, described via partial waves in e.g. \cite{Bautista:2021wfy}, when a NUT
charge $N=-iM$ is included. The amplitude vanishes at $\omega=-p$. This agrees with the self-dual gravity prediction, in particular, the scalar result of \cite{Adamo:2023fbj}. On the other hand setting $\omega=p$, we obtain
\begin{equation}
    \mathcal{\mathcal{A}}_{\omega=p} =  \frac{2M e^{ia\cdot k}}{\sin^{2-i4M\omega}\theta/2}\frac{1}{\Gamma(1+i4M\omega)}.
\end{equation}
The analytic structure of this amplitude reveals an interesting feature. On one hand, there is the usual branch cut along $t\propto \sin^2 \theta/2$ without mass gap. On the other hand, it features \textit{zeros} in $s\propto 2M\omega$, precisely located along the imaginary axis $\omega = \frac{in}{4M}$. Zeroes of scattering amplitudes usually emerge in the unphysical sheet as reflections of poles, suggesting the analytic continuation done here requires further care.

\subsection{General Helicity }

The Teukolsky equation in a rotating black hole \eqref{eq:sd1xa} was
computed in \cite{Bini:2003sy} using the method of spin coefficients.
For instance, for $s=\pm2$ the scattering of gravitational waves
is described in terms of the following scalar

\begin{align}
\Psi_{s=-2} & =(r-iN+a\cos\theta)^{4}\psi_{4}, \\
\Psi_{s=2} & =\psi_{0},\nonumber 
\end{align}
where $\psi_{4},\psi_{0}$ are the Weyl scalars for the linearized
metric perturbation. Analogous definitions hold for $\Psi_{s}$ with
$s=0,\pm1/2,\pm1,\pm3/2,\pm2$\footnote{Due to the peeling theorem we expect $\Psi_{s}$ has $r^{-1}$ and
$r^{-2s-1}$ decaying components as $r\to\infty$.}. It was shown that the master Teukolsky equation can be written in
the compact form
\begin{equation}
T[\Psi_{s}]=(\nabla^{\mu}+s\Gamma^{\mu})(\nabla_{\mu}+s\Gamma_{\mu})\Psi_{s}-4s^{2}\bar{\psi}_{2}\Psi_{s}=0\label{eq:mtq}
\end{equation}
using covariant derivatives w.r.t the background metric \eqref{eq:sd1xa},
a connection vector $\Gamma^{\mu}$ (that can be found in \cite{Bini:2003sy}
and the only non-vanishing Weyl scalar of the background, $\bar{\psi}_{2}$
(in Lorentzian signature):
\begin{equation}
\bar{\psi}_{2}:=\frac{M-iN}{(r-iN+a\cos\theta)^{3}}\,.\label{eq:}
\end{equation}
We now inspect this equation for the SD black hole, which has $\bar{\psi}_{2}=0$.
It turns out that when setting $M=iN$ one can rewrite (\ref{eq:mtq})
in a remarkably simple form
\begin{equation}
\sqrt{1+\frac{2M}{r}}T[\Psi_{s}]=(\partial^{\mu}+\mathcal{A}^{s\mu})(\partial_{\mu}+\mathcal{A}_{\mu}^{s})\Psi_{s}=0 ,
\end{equation}
which is an equation of a charged scalar in flat space! Using the
coordinates $(t,r,\theta,\phi)$ as defined in previous section the
potential reads
\begin{equation}
\mathcal{A}^{s}=i(q-s)\left[\frac{dt}{r}-\cos\theta d\phi\right]-s\,\,d\log r=i(q-s)A-s\,\,d\log r\,.
\end{equation}
The first term is simply the dyon potential $\eqref{eq:dp}$, where
the dyon charge $q=2M\omega$ adds to the spin $s$ of the wave. This
is expected classically since $q$ and $s$ contribute to the angular
momentum in the same direction, see \cite{Kol:2021jjc}. The second term
is a pure gauge term which can be removed by 
\begin{equation}
\Psi_{s}=r^{s}\Phi_{s}\,.
\end{equation}
The resulting equation is (define $\bar{q}=q-s$)
\begin{align*}
\left[(p_{i}+i\bar{q}A_{i})^{2}-\frac{\bar{q}^{2}}{r^{2}}\right]\Phi_{s}-2\frac{\omega\bar{q}}{r}\Phi_{s} & =\omega^{2}\Phi_{s}\,,
\end{align*}
which is equivalent to the Schrodinger problem (\ref{eq:scwa}) in
Lorentzian signature, with the replacement $q\to\bar{q}$ and $p\to\omega$.
However, we expect the scattering amplitude contained in $\Phi_{s}$
to have a different fall off than $1/r$. We will leave this analysis
for future work.

\section{Discussion}

In this work we have computed classical orbits and scattering amplitudes for the self dual black hole. Throughout section \ref{sec:geo} we outlined some methods to find the corrections for $N\neq M$, with views towards the astrophysical case $N=0$. Even though the phenomenology and plausible observation of a non-zero NUT charge has a long history (see e.g. \cite{Wu2023,Zhang2021}  for recent discussions on its photon ring), in this work it is regarded as a mechanism to deform an exact integrable system.

The case $N=M$ is interesting on its own, as it correspond to the self-dual gravitational theory. Indeed, our all-PM results for the wavefunction lie half-way between the strict probe limit (geodesics) and the two-body problem. For instance, it is known that the solution to the massless wave equation (i.e. the Teukolsky equation) controls the first correction of the probe limit \cite{Bautista:2021wfy}, closely related to the so-called first self-force approximation \cite{Cheung:2024byb}. It is expected that the integrability derived from the geodesics then persists at this order.

Let us attempt to refine the problem. Consider the following setup: Two bodies of masses $M_1,M_2$ endowed with NUT charges $N_1 ,N_2$ undergo bounded orbits in a classical two-body problem. We ask:

\vspace{1em}

\emph{\textit{How does the perihelion precession depend on the couplings $g_{\pm}=M_1N_2\pm M_2 N_1$? Does it vanish in the self-dual case?}}

\vspace{1em}

Several pieces of evidence suggests the answer to the second question is affirmative. First, the vanishing of the precession in the probe limit to all PM orders and its relation to the scattering amplitude. Second, beyond the probe limit, a similar question was asked for half BPS states in $\mathcal{N}=8$ supergravity in \cite{Caron-Huot:2018ape}. Following their logic, one would seek to compute the 1-loop amplitude exchanging two gravitons with different helicities. But the self-dual theory only has one helicity and therefore it is expected that such correction vanishes (similar reasoning would apply to higher loops). This is consistent with the absence of the $\frac{1}{\sqrt{k^2}}$ term pointed out in section \ref{sec:comps}. Last but not least, the existence of non-perturbative, \textit{static}, multicenter solutions in the self-dual theory \cite{Gibbons:1978tef} suggests integrability for the many body problem itself.

\section*{Acknowledgements}
We thank A. Herderschee, M. Ivanov, D. Kapec, J. Maldacena, R. Monteiro and A. Sharma for useful discussions. A.G. is supported by DOE-SC0009988, `Problems in Theoretical Physics', and EPSRC grant no EP/R014604. He further wishes to acknowledge support from the Harvard Society of Fellows, as well as the Isaac Newton Institute, Cambridge, for support and hospitality during the programme `Twistor Theory' where work on this paper was undertaken. U.K. is supported by the Center
for Mathematical Sciences and Applications at Harvard
University. H. T. acknowledges the support from the Harvard's Purcell Fellowship and the Joseph John Fay III and Phuong-Mai Fay Vietnam and America Scholarship.

\bibliographystyle{utphys}
\bibliography{reference}

\end{document}